\documentclass[fleqn,12pt,twoside]{article}
\usepackage[headings]{espcrc1}

\readRCS
$Id: espcrc1.tex,v 1.2 2004/02/24 11:22:11 spepping Exp $
\usepackage{graphicx}
\usepackage[figuresright]{rotating}

\newcommand{\AmS}{{\protect\the\textfont2
  A\kern-.1667em\lower.5ex\hbox{M}\kern-.125emS}}

\title{Quark Deconfinement inside Compact Stars and \\
Gamma Ray Bursts Inner Engine}

\author{Alessandro Drago, Giuseppe Pagliara and Irene Parenti
\address[MCSD]{Dipartimento di Fisica - Universit\`a di Ferrara and
INFN Sez. di Ferrara\\
Via Saragat, 1 - 44100 Ferrara - Italy\\
email: drago@fe.infn.it}}
       
\begin{document}

\maketitle

\begin{abstract}
The temporal structure of Gamma Ray Bursts can be interpreted 
assuming as a inner engine a neutron star which undergoes a 
progressive compactification via production of strangeness
(hyperons and kaons) and quarks.
We will propose a tentative identification of various emission
periods of the burst with specific structural changes of the star. 
Each of these
modifications of the composition of the compact star takes place as
a deflagration and not as a detonation, so the energy released
in the transition goes mainly into heat and not into a mechanical
wave. This is important in order to avoid an excessive baryonic contamination
of the region surrounding the compact star. In this way a ultrarelativistic
plasma of electron-positron pairs and of photons can be obtained,
powering the Gamma Ray Burst.

\end{abstract}

\section{Introduction}
In last years the data collected by various X-ray satellites and notably by
Beppo-SAX and by Swift have clarified many crucial features of 
the Gamma-Ray Bursts (GRBs) so that a rather precise phenomenological
description of the explosion can now be drawn.
The light curve can be separated roughly in four emission periods,
although some of these features can often be absent in a specific burst
(for a recent review see e.g. \cite{meszaros}).

1) Several bursts present a {\it precursor}, namely a small signal containing
only a tiny fraction of the total energy of the burst, which anticipates the
main event by tens or even hundreds of seconds. The exact fraction of GRBs
showing a clear evidence of a precursor is difficult to estimate due
to the smallness of the signal. It has been speculated
by Lazzati \cite{lazzati} that at least 20$\%$ of the GRBs display a precursor
and this fraction could still be an underestimate due to the difficulty of 
detecting precursors anticipating the main event by only a few seconds.
The duration of the precursor is typically of a few seconds.

2) The main event corresponds to the emission with the highest luminosity
and is present in the highest energy band of the emission spectrum.
The duration of the main event can vary from few seconds (here we are discussing
only long bursts, having durations longer than roughly 2 s) up to hundreds of seconds.
As we will show, it is possible to divide the main event into active periods
whose duration can be related to the activity of the so-called inner engine,
which is the source of the energy of the burst. The active periods are 
separated by quiescent times.
The high energy photons emitted during the main event are probably associated with 
internal shocks of the ultrarelativistic jet produced by the inner engine. 
In this scenario, shells
with different velocities collide among themselves producing the 
detected signal.

3) Swift satellite has recently provided a strong indication that a large fraction
of GRBs, after the main event and an initial drop in luminosity, displays a plateau 
in which the luminosity drops much less rapidly. Inside the plateau some flares
can also be present. The luminosity of the plateau is much smaller than that of 
the main event, but its duration can be much longer, order of thousands of
seconds, so that the total energy released can be comparable to the energy released
during the main event. 
The plateau can be originated by a residual activity of the
inner engine, although other interpretations are possible.

4) At last the luminosity drops steadily and the so-called afterglow begins.
This final prolongated emission is explained in terms of a progressive expansion
and slow-down of the ultrarelativistic jet, which interacts with the 
interstellar medium emitting photons. This last part of the GRB emission
is now the less controversial.

In this contribution we concentrate on the main event and we provide
some suggestions concerning the precursor.

\section{Quiescent times}

\begin{figure}[t]
\begin{center}
\includegraphics*[50,200][400,300]{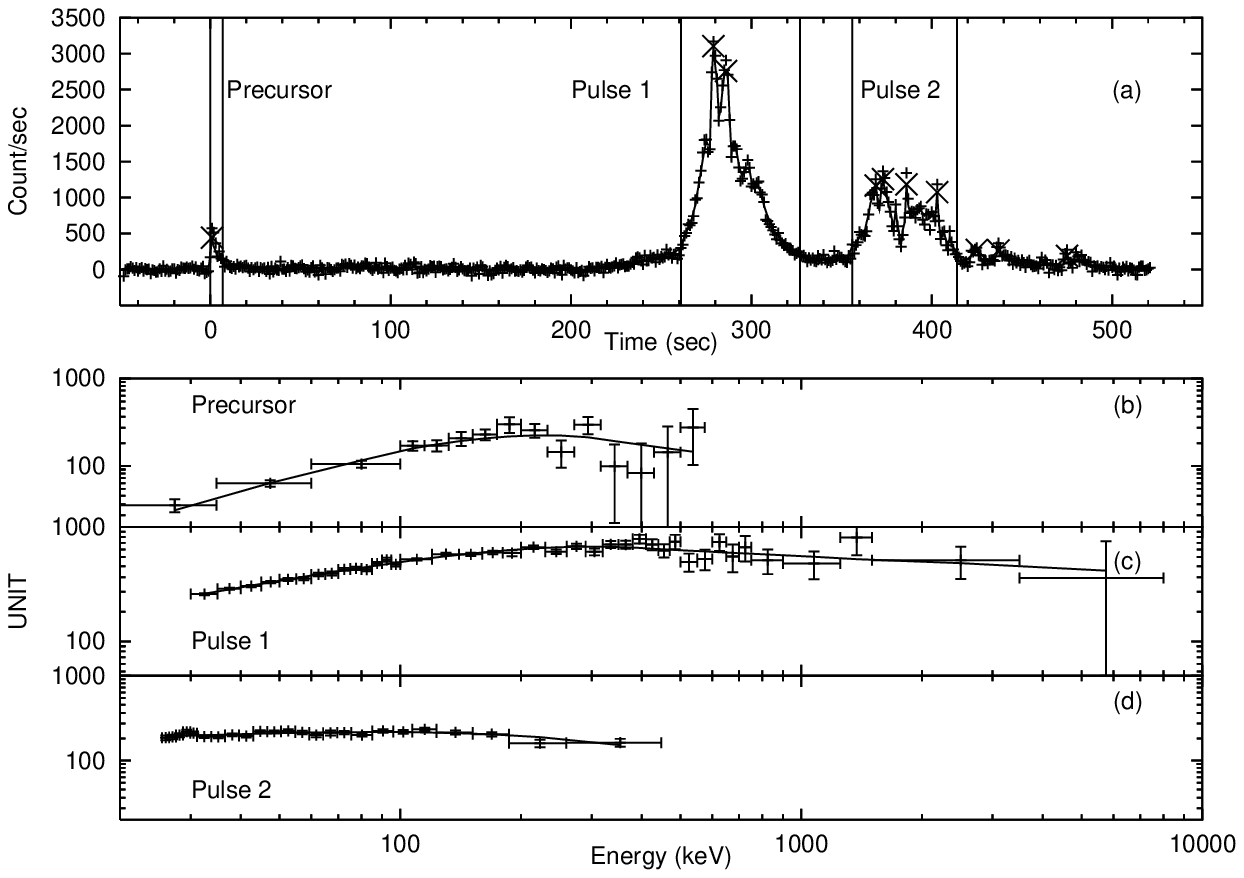}
\end{center}
\parbox{16cm}{
\caption{Example of light curve of a GRB. A precursor
anticipating the main event by 260 s is visible. The main emission
is divided in two active periods separated by a quiescent time
lasting $\sim$ 30 s. From \cite{mcbreen}.}
\label{mcbreen}
}
\end{figure}

\begin{figure}[t]
\begin{center}
\includegraphics[scale=0.5]{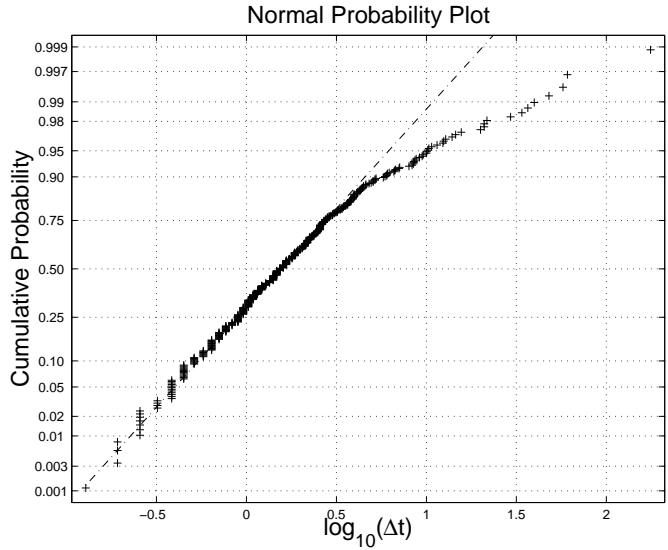}
\end{center}
\parbox{16cm}{
\caption{Cumulative probability distribution of the time intervals
$\Delta t$ between pulses, compared to a best-fit log-normal distribution.
From \cite{np}.}
\label{npfig}
}
\end{figure}

In Fig.~\ref{mcbreen} we show an example of GRB in which precursor and 
quiescent times are clearly visible. The origin of the quiescent times is
still debated. Nakar and Piran \cite{np} suggested on a statistical
basis that the time intervals
during which the GRB shows no activity have a different origin 
than the time intervals
separating peaks within an active period. Fig.~\ref{npfig} 
clearly indicates that 
the number of long quiescent times exceeds a stochastic lognormal distribution.

\begin{figure}[tb]
\begin{center}
\includegraphics[scale=0.6]{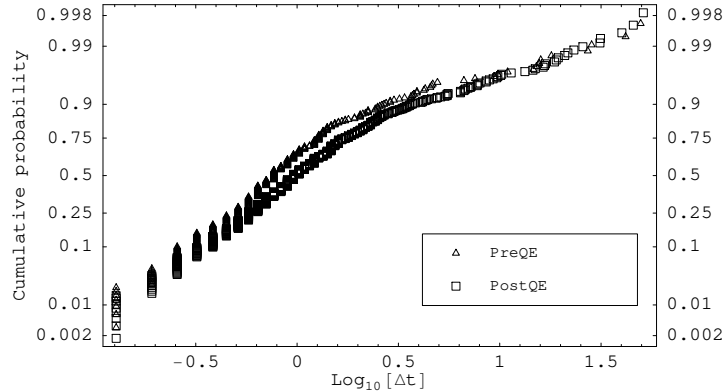}
\end{center}
\parbox{16cm}{
\caption{Cumulative probability distribution of the time intervals
within the PreQuiescent and the PostQuiescent Emission. The two
distributions have a high probability to be equal.}
\label{terza}
}
\end{figure}

\begin{figure}[tb]
\begin{center}
\includegraphics[scale=0.5]{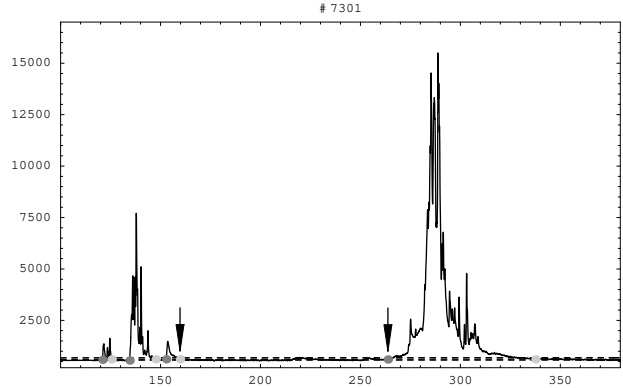}
\end{center}
\parbox{16cm}{
\caption{Example of a GRB having a long quiescent time ($\sim$ 100 s).
The durations of the two active periods ($\sim$ 40 s for PreQE and $\sim$ 80 s for
PostQE) are much shorter than the total duration of the burst ($\sim$ 220 s).}
\label{burst}
}
\end{figure}

We have recently investigated the structure of the 
pre-quiescent and of the post-quiescent emission \cite{pagliara}, 
showing that they share the same
micro-structure (see Fig.~\ref{terza}) and also the same emission power and 
spectral index. Therefore both emissions are generated by a same mechanism
which repeats after a quiescent time. It is therefore rather natural
to interpret this result as due to different activity periods of the inner engine,
during which most of the energy is injected into the fireball. These active periods
are separated by quiescent times during which the inner engine remains dormant.
The advantage of this interpretation is that it reduces the energy request
on the inner engine, the alternative interpretation being that the inner engine
remains active and injects energy also during the quiescent times.
Moreover, in the latter scenario special conditions on the shells velocity
have to be imposed in order to
explain why the emission is strongly suppressed although the inner
engine remains active.
It is possible to show that all GRBs of the BATSE catalogue can be explained
by assuming two active periods (which in many cases merge and are therefore 
not distinguishable). After taking into
account the cosmological correction on time intervals
$t\rightarrow t/(1+z)$ with $z \sim 2$ for BATSE, 
the duration of each active period does not exceed $\sim 30$ s.
In Fig.~\ref{burst} we show an example of 
GRB where the duration of the active periods is significantly
smaller than the total time interval between the beginning and the end of the main
event.

\section{Hadrons to quarks conversion: detonation or deflagration?}
It has been proposed several times that the transition from a star
containing only hadrons to a star composed, at least in part, of 
deconfined quarks can release enough energy to power a GRB 
\cite{cheng,wang,ouyed,apj,haensel}.
A crucial question concerns the way in which the conversion takes place,
either via a detonation or a deflagration. It has been shown in the past that
the mechanical wave associated to a detonation would expel a relatively large
amount of baryon from the star surface \cite{fryer}. In the case 
of a detonation the region in which the 
electron-photon plasma forms (via neutrino-antineutrino annihilation near the
surface of the compact star) would be contaminated by the baryonic load and it
would be impossible to accelerate the plasma up to the enormous Lorentz factors
needed to explain the GRBs.

\begin{figure}[tb]
\begin{center}
\includegraphics[scale=0.4]{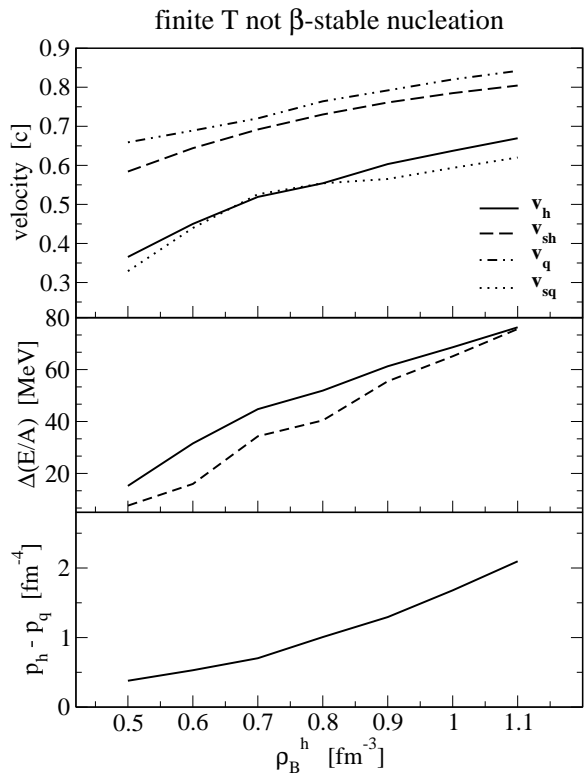}
\end{center}
\parbox{16cm}{
\caption{Upper panel: velocity of hadronic phase $v_h$, of the burned phase
$v_q$ and corresponding sound velocities $v_{sh}$ and $v_{sq}$, all in
units of the velocity of light and in the front frame.  Center panel:
energy difference between the two phases (in the hadron phase rest
frame).  The dashed and the solid lines correspond respectively to the first and to
the second iteration in the solution of the fluidodynamics equations.
Lower panel: pressure difference between the uncombusted and the combusted
phase.  Here the combusted phase is obtained using $B^{1/4}=170$ MeV,
temperatures from 5 to 40 MeV (as estimated from the solid line in the
central panel) and it is not $\beta$-stable.}
\label{velocita}
}
\end{figure}

We have shown in a recent paper \cite{irene}
that the process of conversion always takes place
through a deflagration and not a detonation. In principle the problem of
classifying the conversion process can be solved by comparing the velocity of
the conversion front to the velocity of sound in the unburned phase.
If the velocity of the front it subsonic the process is a deflagration.
The velocity of conversion can be estimated in first approximation
through energy-momentum and
baryon flux conservations through the front. In Fig.~\ref{velocita}
we show the result of such a calculation, indicating that the 
conversion goes through a deflagration with an unstable front.
The instability of the front can be deduced by observing that
the velocity of sound in the burned phase is smaller than
the velocity of that phase in the front frame. 
The temperature released in the
conversion can also be estimated using the thermodynamics first principle.

\begin{figure}[tb]
\begin{center}
\includegraphics*[scale=0.4]{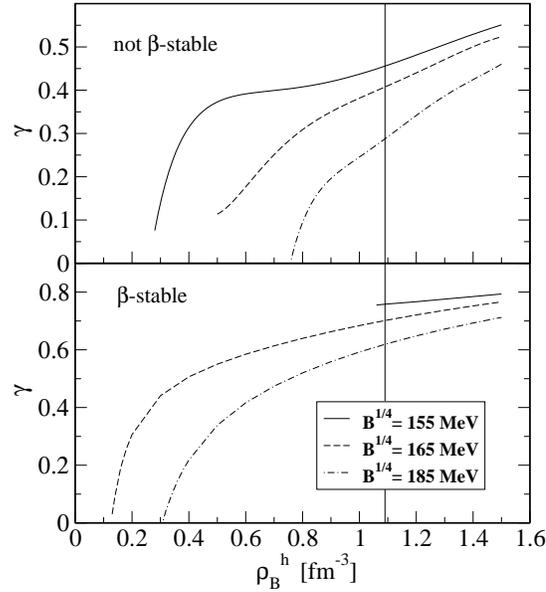}
\end{center}
\parbox{16cm}{
\caption{The $\gamma$-factor entering the fractal dimension of the conversion front.
See Sec.~3.}
\label{landau}
}
\end{figure}

The problem of computing the actual conversion velocity
is anyway more complicated due to fluidodynamical and 
convective instabilities.
Fluidodynamical instabilities are associated with the possibility
of the front to form wrinkles. In this way the surface area increases and the
conversion can accelerate respect to the laminar velocity $v_{lam}$
\cite{blinnikov}. In the absence of new dimensional
scales between the minimal dimension $l_{\mathrm{min}}$ and the maximal dimension $l_{\mathrm{min}}$
of the wrinkle,
the effective velocity is given by the expression
\begin{equation}
v_{\mathrm{eff}}=v_{\mathrm{lam}} \left ( \frac {l_{\mathrm{max}}}{l_{\mathrm{min}}}\right ) ^{D-2}\, .
\end{equation}
Here $D$ is the fractal dimension of the surface of the front and it can be estimated
as $D=2+D_0 \gamma^2$, where $D_0\sim 0.6$ and $\gamma=1-\rho_b/\rho_u$.
In this analysis a crucial role is played by neutrino trapping which does not allow
the system to reach $\beta$-equilibrium on the same timescale of the conversion
process. Taking into account this delay of the weak processes, then $\gamma\le 0.45$
at all densities. The effect of neutrino trapping is displayed in 
Fig.~\ref{landau}.
Our numerical analysis shows that, although the effective velocity can be 
significantly enhanced respect to the laminar velocity, it is unlikely that 
$v_{\mathrm{eff}}$ exceeds the speed of sound and therefore the process remains a deflagration.

\begin{figure}[tb]
\begin{center}
\includegraphics*[scale=0.33]{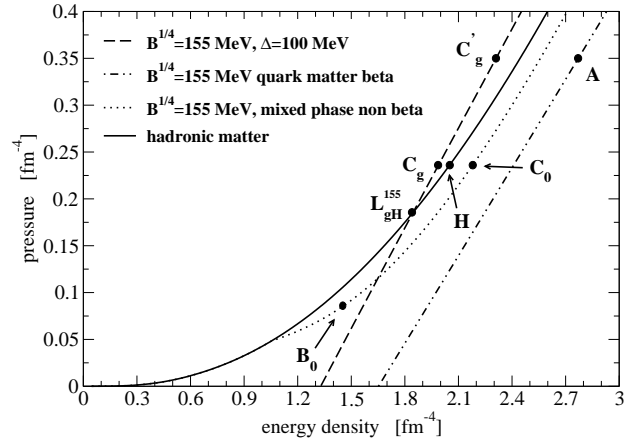}
\end{center}
\parbox{16cm}{
\caption{Scheme for convection: H represents the drop of hadronic matter just
before deconfinement, B$_0$ represents the drop of newly formed QM, C
stays for the drop of QM after pressure equilibration and L indicates
the end point of the convective layer. Finally A represents a drop of
ungapped quark matter before its transition to CFL phase. From \cite{irene}.}
\label{convection}
}
\end{figure}

Convective instability can also instaure, 
because in a regime of strong
deflagration the energy density of the newly formed phase is
smaller than the energy density of the old phase. On the other hand,
in a high density system in which relativistic corrections are important
the new phase forms at a pressure smaller than the pressure of the old phase
(here matter is not yet at equilibrium,
which is reached only after a delay). Due to this, when the drop of new phase
enters the old phase pushed by the gravitational gradient, its pressure
rapidly re-equilibrates and its energy density changes accordingly.
Quasi-Ledoux convection develops only if the energy density of the new phase
remains smaller than that of the old phase {\it after} the pressure has equilibrated. 
In Fig.~\ref{convection} we show a typical analysis which allows to decide if
convection can take place and, in case, which is the convective layer.

Summarizing, the results of our analysis are the followings:
\begin{itemize}
\item
the conversion always takes place as a strong deflagration and never as a detonation
\item
fluidodynamical instabilities are present and they significantly
increase the conversion velocity but, in realistic cases, 
the conversion process does not transform from a deflagration to a detonation
\item
convection can develop in specific cases, in particular it takes place if hyperons are present or if
diquark condensate does form.
\end{itemize}

\section{Structural modifications of the compact star and light curves of the GRBs}
We can combine the information provided in the previous sections and formulate
a model for the GRBs based on the following scheme:
\begin{itemize}
\item
a compact star forms after a Supernova explosion. The explosion can be 
entirely successful or marginally failed, so that in both cases the mass-fallback
is moderate (fraction of a solar mass);
\item
after a delay, varying from seconds to years and dependent 
on the mass of the compact star and on the mass accretion rate,
the star starts readjusting its internal structure. The first event could be
associated with the formation of kaon condensation (or of hyperons if it goes
through a first order transition \cite{bielich}). This first structural
modification could be relatively small, involving only a modification
of the central region of the star, but the presence of strangeness can
trigger the instability respect to the formation of strange quark matter.
The precursors could be due to this process;
\item
the compact star is now metastable respect to the formation of quark matter
(if deconfinement at finite density takes place as a first order transition)
and after a short delay the formation of deconfined quarks takes place as a 
deflagration. A hot compact star remains, and it cools-down through neutrino-antineutrino
emission;
\item
many calculations indicate that Color-Flavor-Locked (CFL) quark matter is the
most stable configuration at large density. On the other hand the transition
from normal quarks to CFL matter can take place as a first order if the
leptonic content of the newly formed normal quark matter phase is not too small
\cite{ruster}
(its initial leptonic content equals that of the hadronic compact star).
In that way, after a short delay (quiescent time) a second transition can take
place inside the compact star, due to the formation of superconducting quarks.
Energy is again released, and a hot and more compact stellar object is now
formed, which again starts cooling via neutrino emission;
\item
the neutrino-antineutrino emitted by the compact star can annihilate near the surface
with an efficiency of order percent. The energy deposited in the electron-positron-gamma
plasma can be large enough to power a GRB.

\end{itemize}

\noindent
It is interesting to compare the scheme here proposed
to the hypernova-collapsar model. In that model the GRB can be associated
with a SN explosion which has to be strictly simultaneous with the GRB. In
the quark deconfinement model the two events can be temporally separated, with the
SN preceding the GRB by a delay which can vary from minutes to years.
Arguments in favor of a two-steps mechanism have been discussed in the
literature \cite{dermer}.

\smallskip
\noindent
We would like to thank Andrea Lavagno for a longstanding collaboration.
Most of the results here discussed have been obtained working together.

\end{document}